\documentclass[12pt]{article}

\usepackage{epsfig}

\usepackage{bbm}
\usepackage{amssymb,amsmath}

\newcommand{\be}{\begin{equation}}
\newcommand{\ee}{\end{equation}}
\newcommand{\ba}{\begin{eqnarray}}
\newcommand{\ea}{\end{eqnarray}}
\newcommand{\no}{\nonumber\\}

\newcommand{\mnu}{\mathcal{M}_\nu}

\newcommand{\zz}{\mathbbm{Z}_2}

\begin{document}

\title{\normalsize \hfill UWThPh-2008-19 \\[8mm]
\LARGE Tri-bimaximal lepton mixing from symmetry only}

\author{
W.~Grimus$^{(1)}$\thanks{E-mail: walter.grimus@univie.ac.at}
\ and
L.~Lavoura$^{(2)}$\thanks{E-mail: balio@cftp.ist.utl.pt}
\\*[3mm]
$^{(1)}$ \small
University of Vienna, Faculty of Physics \\
\small
Boltzmanngasse 5, A--1090 Vienna, Austria
\\*[2mm]
$^{(2)}$ \small
Technical University of Lisbon,
Centre for Theoretical Particle Physics \\
\small Instituto Superior T\'ecnico, 1049-001 Lisbon, Portugal
}

\date{24 February 2009}

\maketitle

\begin{abstract}
We construct a model for tri-bimaximal
lepton mixing which employs only family symmetries and their soft breaking;
neither vacuum alignment nor supersymmetry,
extra dimensions,
or non-renormalizable terms are used in our model.
It is an extension of the Standard Model
making use of the seesaw mechanism
with five right-handed neutrino singlets.
The scalar sector comprises four Higgs doublets
and one complex gauge singlet.
The horizontal symmetry of our model
is based on the permutation group $S_3$ of the lepton families
together with the three family lepton numbers---united this constitutes
a symmetry group $\Delta(6\infty^2)$.
The model makes no predictions for the neutrino masses. 
\end{abstract}

\newpage

\section{Introduction}

Lepton mixing and non-zero neutrino masses
are now established facts---for reviews
and for the latest fits see~\cite{results}.
The mixing angles in the lepton mixing matrix $U$
have values quite different from those of quark mixing.
The phenomenological hypothesis that
\be
U = U_\mathrm{HPS} \equiv \left( \begin{array}{ccc}
2/\sqrt{6} & 1/\sqrt{3} & 0 \\ 
-1/\sqrt{6} & 1/\sqrt{3} & -1/\sqrt{2} \\ 
-1/\sqrt{6} & 1/\sqrt{3} & 1/\sqrt{2}
\end{array} \right)
\label{HPS}
\ee
has been put forward by Harrison,
Perkins and Scott (HPS) in 2002~\cite{HPS}.
At present,
all the experimental data are still compatible
with this simple ``tri-bimaximal'' mixing \textit{Ansatz}.

The hypothesis~(\ref{HPS}) has stimulated model building
and the search for family symmetries
which might lead to $U = U_\mathrm{HPS}$ in a natural way.
While it is not difficult to simultaneously obtain $U_{e3}=0$
and maximal atmospheric-neutrino mixing~\cite{GL01},
generating a solar mixing angle
$\theta_{12} = \arcsin{\left( 1 \left/ \sqrt{3} \right. \right)}$
is highly non-trivial and in general necessitates complicated models.
In those models one often finds
several scalar multiplets of the horizontal-symmetry group
with vacuum expectation values (VEVs)
aligned in a special way.
To explain this peculiar alignment of VEVs
one may have recourse to special scalar potentials,
stabilized with the help of supersymmetry---see
for instance~\cite{aligned,prescient,merlo}---or 
to extra-dimensional models~\cite{extradim}.

In two previous papers~\cite{trimax,tricp} 
we have enforced trimaximal mixing---which is a weaker hypothesis
than tri-bimaximal mixing---through a model.
We now show that,
with very little extra effort,
one can also achieve tri-bimaximal mixing along the same lines.
In the model that we shall present here
neither VEV alignment nor supersymmetry,
non-renormalizable terms,
or extra dimensions are required for obtaining $U = U_\mathrm{HPS}$.
Besides enlarging the scalar sector of the Standard Model (SM)
by several Higgs doublets and one gauge singlet,
our model uses the seesaw mechanism~\cite{seesaw}
with \emph{more than three} right-handed neutrino singlets,
but in such a way that the additional right-handed neutrinos
do not have Yukawa couplings to the Higgs doublets;
then these additional right-handed neutrinos---in
the present case there are two of them---can be
exploited for imposing the desired mixing properties.\footnote{This idea
had already been previously used by us for tri-bimaximal mixing,
but in that case we still needed VEV alignment
and made use of supersymmetry~\cite{prescient}.}
In our model lepton mixing originates solely
in the Majorana mass matrix $M_R$ of the right-handed neutrino singlets, 
and the number of independent Yukawa coupling constants of the Higgs doublets
is an absolute minimum---only two.

This paper is organized as follows.
The model is presented in section~\ref{model}.
Variations on the symmetries of the model,
and their connection to the renormalization-group evolution (RGE)
of the light-neutrino mass matrix $\mnu$,
are investigated in section~\ref{variations}.
The conclusions are presented in section~\ref{concl}.
An appendix contains details
of the computation of the $3 \times 3$ matrix $\mnu$
out of the $5 \times 5$ matrix $M_R$.

\section{The model}
\label{model}

\subsection{Fields and symmetries}
\label{fs}

Our model is based on the SM gauge group $SU(2) \times U(1)$.
The lepton sector\footnote{We neglect the quark sector,
which is immaterial for our purposes.}
consists of three left-handed $SU(2)$ doublets
$D_{\alpha L} = \left( \nu_{\alpha L}, \alpha_L \right)^T$
($\alpha = e, \mu, \tau$),
three right-handed charged-lepton $SU(2)$ singlets $\alpha_R$,
and \emph{five} right-handed $SU(2) \times U(1)$ singlet neutrinos
$\nu_{\alpha R}$,
$\nu_{\ell R}$ ($\ell = 1, 2$).
The scalar sector consists of
one complex gauge singlet $\chi$ with zero electric charge 
and four Higgs doublets
$\phi_\alpha = \left( \phi_\alpha^+, \phi_\alpha^0 \right)^T$,
$\phi_0 = \left( \phi_0^+, \phi_0^0 \right)^T$.

The family symmetries of the model are the following:
\begin{itemize}
\item Three $U(1)$ symmetries
associated with the family lepton numbers $L_\alpha$,
\be
\label{U1}
U(1)_{L_\alpha}: \quad D_{\alpha L} \to e^{i \psi_\alpha} D_{\alpha L}, \
\alpha_R \to e^{i \psi_\alpha} \alpha_R, \
\nu_{\alpha R} \to e^{i \psi_\alpha} \nu_{\alpha R},
\quad \psi_\alpha \in \left[ 0, 2 \pi \right[.
\ee
The $U(1)_{L_\alpha}$
are supposed to be \emph{softly} broken at high energy,
\textit{i.e.}~at the seesaw scale~\cite{GL01,GL00},
by \emph{dimension-three} terms of the types
$\nu_{\alpha L}^T C^{-1} \nu_{\beta L}$,
$\nu_{\alpha L}^T C^{-1} \nu_{\ell L}$
($C$ is the Dirac--Pauli charge-conjugation matrix).
\item The $S_3$ permutation symmetry of the $e, \mu, \tau$ indices.
We view this permutation symmetry
as being generated by two non-commuting transformations:
\begin{itemize}
\item The cyclic transformation
\be
\label{Cemutau}
C_{e \mu \tau}: \quad \left\{ \begin{array}{l}
D_{e L} \to D_{\mu L} \to D_{\tau L} \to D_{e L}, \\
e_R \to \mu_R \to \tau_R \to e_R, \\
\nu_{e R} \to \nu_{\mu R} \to \nu_{\tau R} \to \nu_{e R}, \\
\phi_e \to \phi_\mu \to \phi_\tau \to \phi_e, \\
\nu_{1R} \to \omega \nu_{1R}, \ \nu_{2R} \to \omega^2 \nu_{2R}, \\
\chi \to \omega \chi, \ \chi^\ast \to \omega^2 \chi^\ast,
\end{array} \right.
\ee
where $\omega \equiv \exp{\left( 2 i \pi / 3 \right)}$
is the cubic root of unity with the properties 
$\omega^2 = \omega^\ast$ and $1 + \omega + \omega^2 = 0$.
\item The $\mu$--$\tau$ interchange~\cite{GL01}
\be
\label{Imutau}
I_{\mu \tau}: \quad \left\{ \begin{array}{l}
D_{\mu L} \leftrightarrow D_{\tau L}, \\
\mu_R \leftrightarrow \tau_R, \\
\nu_{\mu R} \leftrightarrow \nu_{\tau R}, \\
\phi_\mu \leftrightarrow \phi_\tau, \\
\nu_{1R} \leftrightarrow \nu_{2R}, \\
\chi \leftrightarrow \chi^\ast.
\end{array} \right.
\ee
\end{itemize}
It is clear that the fields with $\alpha$ indices
form triplet \emph{reducible} representations of $S_3$,
while
\[
\left( \begin{array}{c}
\nu_{1R} \\ \nu_{2R}
\end{array} \right),
\quad
\left( \begin{array}{c}
\chi \\ \chi^\ast
\end{array} \right)
\]
transform under $S_3$ according to the complex version
of the doublet irreducible representation,
previously used for instance in~\cite{S3Z2}.\footnote{If one wishes
one may separate $\chi$ into its real and imaginary parts,
which transform under $S_3$
according to the real version of the doublet irreducible representation.}
The cyclic transformation $C_{e \mu \tau}$
is softly broken by \emph{dimension-two} and \emph{dimension-one} terms
in the scalar potential,
but it is \emph{preserved} by all the \emph{dimension-three}
(and,
of course,
dimension-four)
terms in the Lagrangian.
The symmetry $I_{\mu \tau}$ is \emph{not} allowed to be softly broken.
The VEV $v_\chi \equiv \left\langle \chi \right\rangle_0$
breaks $C_{e \mu \tau}$ spontaneously,
but it preserves $I_{\mu \tau}$ because it is \emph{real};
this is a consequence of the $I_{\mu \tau}$-invariance
of the scalar potential,
as will be shown in subsection~\ref{potential}.
At low energy,
both $C_{e \mu \tau}$ and $I_{\mu\tau}$ are spontaneously broken
because all three vacuum expectation values (VEVs)
$v_\alpha \equiv \left\langle \phi_\alpha^0 \right\rangle_0$
are different (see below). 
\item Three $\mathbbm{Z}_2$ symmetries~\cite{prescient,nasri}
\be
\label{Z2}
\mathbbm{Z}_2^{(\alpha)}: \quad \alpha_R \to - \alpha_R, \
\phi_\alpha \to - \phi_\alpha,
\ee
for $\alpha = e, \mu, \tau$.
The $\mathbbm{Z}_2^{(\alpha)}$
are supposed to be \emph{softly} broken at low energy,
\textit{i.e.}~at the electroweak scale,
by \emph{dimension-two} terms of the types
$\phi_\alpha^\dagger \phi_\beta$ ($\alpha \neq \beta$),
$\phi_\alpha^\dagger \phi_0$.
The symmetry $\mathbbm{Z}_2^{(\alpha)}$ is spontaneously broken
when $\phi_\alpha^0$ acquires the non-zero VEV $v_\alpha$.
\end{itemize}

\subsection{Lagrangian and lepton mixing}

The Yukawa Lagrangian has dimension four
and therefore respects all the symmetries of the model.
It is given by
\begin{subequations}
\label{yuwa}
\ba
\mathcal{L}_\mathrm{Yukawa} &=&
- y_1 \sum_{\alpha = e, \mu, \tau } \bar D_{\alpha L} \alpha_R \phi_\alpha
\label{alphar} \\ & &
- y_2 \sum_{\alpha = e, \mu, \tau}
\bar D_{\alpha L} \nu_{\alpha R}
\left( i \tau_2 \phi_0^\ast \right)
\label{nualphar} \\ & &
+ \frac{y_3}{2} \left( \chi \, \nu_{1R}^T C^{-1} \nu_{1R} +
\chi^\ast \, \nu_{2R}^T C^{-1} \nu_{2R} \right)
+ \mathrm{H.c.}
\label{chi}
\ea
\end{subequations}
The symmetries $\mathbbm{Z}_2^{(\alpha)}$ are instrumental
in ensuring that only the doublet $\phi_\alpha$
couples to $\alpha_R$---line~(\ref{alphar})---and that
only the doublet $\phi_0$
couples to the three $\nu_{\alpha R}$---line~(\ref{nualphar}).
The family-lepton-number symmetries $U(1)_{L_\alpha}$ are also important
to enforce Yukawa couplings diagonal in flavour space~\cite{GL01}.
Note that the number of Yukawa coupling constants
of the Higgs doublets is an absolute minimum---just $y_1$ and $y_2$.

Upon spontaneous symmetry breaking (SSB)
the charged leptons acquire masses
$m_\alpha = \left| y_1 v_\alpha \right|$.
Since those three masses are supposed to be all different,
the scalar potential must be rich enough that
the VEVs $v_\alpha$ turn out to be all different.
Also upon SSB the neutrinos acquire,
from line~(\ref{nualphar}),
Dirac mass terms
\be
- \left( \begin{array}{ccccc}
\bar \nu_{eR} & \bar \nu_{\mu R} & \bar \nu_{\tau R} &
\bar \nu_{1R} & \bar \nu_{2R} \end{array} \right) M_D
\left( \begin{array}{c}
\nu_{e L} \\ \nu_{\mu L} \\ \nu_{\tau L} \end{array} \right)
+ \mathrm{H.c.},
\ee
where 
\be
\label{MD}
M_D = \left( \begin{array}{ccc}
a & 0 & 0 \\ 0 & a & 0 \\ 0 & 0 & a \\
0 & 0 & 0 \\ 0 & 0 & 0
\end{array} \right),
\quad a \equiv y_2^\ast v_0,
\quad v_0 \equiv \left\langle \phi_0^0 \right\rangle_0.
\ee

In the Lagrangian there are also bare neutrino Majorana mass terms.
These terms have dimension three and are,
therefore,
allowed to break the family lepton numbers,
but not the permutation symmetry $S_3$.
They are
\begin{subequations}
\ba
\mathcal{L}_\mathrm{Majorana} &=&
\frac{M_0^\ast}{2} \sum_{\alpha = e, \mu, \tau}
\nu_{\alpha R}^T C^{-1} \nu_{\alpha R}
\label{m0} \\ & &
+ M_1^\ast \left(
\nu_{eR}^T C^{-1} \nu_{\mu R}
+ \nu_{\mu R}^T C^{-1} \nu_{\tau R}
+ \nu_{\tau R}^T C^{-1} \nu_{eR}
\right)
\label{m1} \\*[1mm] &&  
+ M_2^\ast \left[ \nu_{1R}^T C^{-1} \left(
\nu_{eR} + \omega \nu_{\mu R} + \omega^2 \nu_{\tau R} \right)
\right. \no & & \left.
+ \nu_{2R}^T C^{-1} \left(
\nu_{eR} + \omega^2 \nu_{\mu R} + \omega \nu_{\tau R} \right)
\right] 
\label{m2} \\*[1mm] & &
+ M_4^\ast \nu_{1R}^T C^{-1} \nu_{2R}
+ \mathrm{H.c.}
\label{LM}
\ea
\label{majorana}
\end{subequations}
Together with line~(\ref{chi}) upon SSB,
$\mathcal{L}_\mathrm{Majorana}$ generates the neutrino Majorana mass terms
\be
- \frac{1}{2}
\left( \begin{array}{ccccc}
\bar \nu_{eR} & \bar \nu_{\mu R} & \bar \nu_{\tau R} &
\bar \nu_{1R} & \bar \nu_{2R} \end{array} \right)
M_R C
\left( \begin{array}{c}
\bar \nu_{eR}^T \\ \bar \nu_{\mu R}^T \\ \bar \nu_{\tau R}^T \\
\bar \nu_{1R}^T \\ \bar \nu_{2R}^T \end{array} \right)
+ \mathrm{H.c.},
\ee
where the symmetric matrix $M_R$ is
\be
M_R = \left( \begin{array}{ccccc}
M_0 & M_1 & M_1 & M_2 & M_2 \\
M_1 & M_0 & M_1 & \omega^2 M_2 & \omega M_2 \\
M_1 & M_1 & M_0 & \omega M_2 & \omega^2 M_2 \\
M_2 & \omega^2 M_2 & \omega M_2 & M_N & M_4 \\
M_2 & \omega M_2 & \omega^2 M_2 & M_4 & M^\prime_N 
\end{array} \right),
\quad M_N \equiv y_3^\ast v_\chi^\ast,
\quad M^\prime_N \equiv y_3^\ast v_\chi.
\ee

We now derive the effective light-neutrino Majorana mass terms
\be
\mathcal{L}_\nu =
\frac{1}{2}
\left( \begin{array}{ccc}
\nu_{eL}^T & \nu_{\mu L}^T & \nu_{\tau L}^T
\end{array} \right)
C^{-1} \mnu
\left( \begin{array}{c}
\nu_{eL} \\ \nu_{\mu L} \\ \nu_{\tau L}
\end{array} \right)
+ \mathrm{H.c.},
\ee
where
\be
\mnu = - M_D^T M_R^{-1} M_D
\ee
according to the seesaw formula~\cite{seesaw}.
Because of the special form of $M_D$ in equation~(\ref{MD}),
only the $3 \times 3$ upper-left submatrix of $M_R^{-1}$ matters.
One finds (for details see appendix~A)
\be\label{mnu}
\mnu = \left( \begin{array}{ccc}
x + y + t &
z + \omega^2 y + \omega t &
z + \omega y + \omega^2 t \\
z + \omega^2 y + \omega t &
x + \omega y + \omega^2 t &
z + y + t \\
z + \omega y + \omega^2 t &
z + y + t &
x + \omega^2 y + \omega t
\end{array} \right).
\ee
Equations~(\ref{y}, \ref{t}) with $M_2 = M_3$ tell us that
\be
\left( y, t \right) \propto \left( M^\prime_N, M_N \right).
\ee
Therefore,
$y / t = v_\chi / v_\chi^\ast$.
We now make the crucial assumption that the VEV $v_\chi$ is real.
This is \emph{not} an unjustified assumption
since it simply corresponds to the conservation of the symmetry $I_{\mu \tau}$
by the VEV of $\chi$.
It follows from this assumption that $t = y$,
hence
\be\label{mnuHPS}
\mnu = \left( \begin{array}{ccc}
x + 2 y &
z - y &
z - y \\
z - y &
x - y &
z + 2 y \\
z - y &
z + 2 y &
x - y
\end{array} \right).
\ee
This is precisely the $\mnu$ corresponding to tri-bimaximal mixing.
Its diagonalization reads
\begin{subequations}
\ba
U_\mathrm{HPS}^T \mnu U_\mathrm{HPS} &=&
\mbox{diag} \left( \mu_1, \mu_2, \mu_3 \right),
\\
\mu_1 &=& x + 3y - z,
\\
\mu_2 &=& x + 2z,
\\
\mu_3 &=& x - 3y - z.
\ea
\end{subequations}
The light-neutrino masses are given by $m_j = | \mu_j |$
($j = 1, 2, 3$).
The matrix $\mnu$ has five parameters,
corresponding to the three neutrino masses and the two Majorana phases,
which are completely free.

\subsection{Scalar potential} \label{potential}

We have demonstrated that our model leads,
under the sole assumption that the VEV $v_\chi$ is real,
to HPS mixing.
In order to check that a real $v_\chi$ is viable,
we proceed to analyze the scalar potential $V$ of the $\phi_m$
($m = 0, e, \mu, \tau$)
and $\chi$.
The potential must respect both
the three symmetries $\mathbbm{Z}_2^{(\alpha)}$
and the permutation symmetry $S_3$,
except for the dimension-two and dimension-one terms,
which are allowed to break softly
both the $\mathbbm{Z}_2^{(\alpha)}$ and $C_{e \mu \tau}$,
but not $I_{\mu \tau}$.
Therefore,
\begin{subequations}
\label{V}
\ba
V &=& \lambda_1 \left[
\left( \phi_e^\dagger \phi_e \right)^2
+ \left( \phi_\mu^\dagger \phi_\mu \right)^2
+ \left( \phi_\tau^\dagger \phi_\tau \right)^2
\right]
+ \lambda_2 \left( \phi_0^\dagger \phi_0 \right)^2
\\ & &
+ \lambda_3 \left(
\phi_e^\dagger \phi_e \, \phi_\mu^\dagger \phi_\mu
+ \phi_\mu^\dagger \phi_\mu \, \phi_\tau^\dagger \phi_\tau
+ \phi_\tau^\dagger \phi_\tau \, \phi_e^\dagger \phi_e
\right)
\\ & &
+ \lambda_4 \, \phi_0^\dagger \phi_0 \left(
\phi_e^\dagger \phi_e
+ \phi_\mu^\dagger \phi_\mu
+ \phi_\tau^\dagger \phi_\tau
\right)
\\ & &
+ \lambda_5 \left(
\phi_e^\dagger \phi_\mu \, \phi_\mu^\dagger \phi_e
+ \phi_\mu^\dagger \phi_\tau \, \phi_\tau^\dagger \phi_\mu
+ \phi_\tau^\dagger \phi_e \, \phi_e^\dagger \phi_\tau
\right)
\\ & &
+ \lambda_6 \, \phi_0^\dagger \left(
\phi_e \phi_e^\dagger
+ \phi_\mu \phi_\mu^\dagger
+ \phi_\tau \phi_\tau^\dagger
\right) \phi_0
\\
& & \label{l7}
+ \lambda_7 \left[
\left( \phi_e^\dagger \phi_\mu \right)^2
+ \left( \phi_\mu^\dagger \phi_\tau \right)^2
+ \left( \phi_\tau^\dagger \phi_e \right)^2
+ \mathrm{H.c.}
\right]
\\
& & \label{l8}
+ \left\{ \lambda_8 \left[
\left( \phi_0^\dagger \phi_e \right)^2
+ \left( \phi_0^\dagger \phi_\mu \right)^2
+ \left( \phi_0^\dagger \phi_\tau \right)^2
\right]
+ \mathrm{H.c.} \right\}
\\
& & \label{l9}
+ \left[ \lambda_9 \left(
\phi_e^\dagger \phi_e
+ \phi_\mu^\dagger \phi_\mu
+ \phi_\tau^\dagger \phi_\tau
\right)
+ \lambda_{10} \phi_0^\dagger \phi_0 \right] \left| \chi \right|^2
\\ & &
+ \lambda_{11} \left| \chi \right|^4
+ \vartheta_1 \left( \chi^3 + {\chi^\ast}^3 \right)
+ \mu_1 \left| \chi \right|^2
+ \mu_2 \left( \chi^2 + {\chi^\ast}^2 \right)
+ \eta \left( \chi + \chi^\ast \right)
\label{crux} \\ & &
+ \lambda_{12} \left[
\chi^2 \left(
\phi_e^\dagger \phi_e
+ \omega^2 \phi_\mu^\dagger \phi_\mu
+ \omega \phi_\tau^\dagger \phi_\tau
\right)
+ {\chi^\ast}^2 \left(
\phi_e^\dagger \phi_e
+ \omega \phi_\mu^\dagger \phi_\mu
+ \omega^2 \phi_\tau^\dagger \phi_\tau
\right)
\right]
\label{l12} \\ & &
+ \vartheta_2 \left[
\chi \left(
\phi_e^\dagger \phi_e
+ \omega \phi_\mu^\dagger \phi_\mu
+ \omega^2 \phi_\tau^\dagger \phi_\tau
\right)
+ \chi^\ast \left(
\phi_e^\dagger \phi_e
+ \omega^2 \phi_\mu^\dagger \phi_\mu
+ \omega \phi_\tau^\dagger \phi_\tau
\right)
\right]
\label{barm} \\ & & \label{phi2}
+ \left( \begin{array}{cccc}
\phi_0^\dagger & \phi_e^\dagger & \phi_\mu^\dagger & \phi_\tau^\dagger
\end{array}\right)
\left( \begin{array}{cccc}
\mu_3 & \mu_9 & \mu_8 & \mu_8 \\
\mu_9^\ast & \mu_4 & \mu_7 & \mu_7 \\
\mu_8^\ast & \mu_7^\ast & \mu_5 & \mu_6 \\
\mu_8^\ast & \mu_7^\ast & \mu_6 & \mu_5
\end{array} \right)
\left( \begin{array}{c}
\phi_0 \\ \phi_e \\ \phi_\mu \\ \phi_\tau
\end{array}\right).
\ea
\end{subequations}
The only parameters in $V$ which may be complex are $\lambda_8$
and $\mu_{7,8,9}$.
Notice the terms $\mu_2$ and $\eta$ in line~(\ref{crux}),
which break $C_{e \mu \tau}$ softly,
and various terms in line~(\ref{phi2})
which break the $\mathbbm{Z}_2^{(\alpha)}$ (and $C_{e \mu \tau}$) softly.
All these terms,
though,
preserve $I_{\mu \tau}$.
The soft breaking of the $\mathbbm{Z}_2^{(\alpha)}$
in line~(\ref{phi2}) is needed in order to prevent
the appearance of Goldstone bosons if $\lambda_7 = \lambda_8 = 0$
(see later).

We want both $v_\chi$ and the mass of $\chi$
to be at the high (seesaw) scale,
while both the $v_m$ and the masses of the $\phi_m$ components
should be at the low (electroweak) scale.
Therefore we must fine-tune $\lambda_{12}$ and $\vartheta_2$
in lines~(\ref{l12}) and~(\ref{barm}),
respectively,
to be extremely small,
lest they pull the masses of the $\phi_\alpha$ components
up to the seesaw scale.\footnote{This fine-tuning is a weak point
of our model,
but most (non-supersymmetric) models with a very high scale
suffer from the same drawback.}
Once $\lambda_{12}$ and $\vartheta_2$ have been tuned to be very small,
the phase of $v_\chi$ becomes determined only by the terms
in line~(\ref{crux}).
It is clear that,
if $\mu_2$ is chosen negative
and the product $\vartheta_1 \eta$ is chosen positive, 
then the minimum of $V$ will be obtained for a real $v_\chi$,
with sign opposite to the one of $\vartheta_1$ and $\eta$~\cite{tricp}.
We have thus shown that there is a range of parameters of the scalar potential
for which the symmetry $I_{\mu \tau}$ is preserved by the seesaw-scale vacuum,
\textit{i.e.}~for which $v_\chi$ is real.

At low scale $I_{\mu \tau}$ is spontaneously broken
by $\left| v_\mu \right| \neq \left| v_\tau \right|$.
Writing
\[
\left( \left| v_\mu \right|, \left| v_ \tau \right| \right)
\propto
\left( \sin{\theta}, \cos{\theta} \right),
\]
and assuming all VEVs and coupling constants to be real,
we verify that the vacuum potential is,
as a function of $\theta$,
of the form
\[
a + b\, \sin^2{2 \theta} + c\, \sin{2 \theta} + d\, \sqrt{1 + \sin{2 \theta}},
\]
where $c \propto \mu_6$ and $d$ stems from the $\mu_{7,8}$ terms.
Is it clear that a vacuum potential of this form
in general leads to a non-trivial value of $\theta$,
which may moreover be very small if both $c$ and $d$
are chosen much smaller than $b > 0$.

\section{Variations on the symmetries and \\ renormalization-group invariance}
\label{variations}

\paragraph{The group structure of the model:}
\emph{All} the symmetries of the model,
and their respective breaking mechanisms,
have been listed in section~\ref{fs},
and in principle it is not necessary to detail the group that they generate.
Still,
elucidating the group structure of the model
may be useful for understanding the terms allowed in the Lagrangian. 
Following for instance the reasoning in~\cite{GL07},
the symmetry group $G$ of our model
may be described as the semidirect product
\be
\label{G}
G = (N \times H) \rtimes S_3,
\ee
where
\begin{description}
\item $N = \zz^{(e)} \times \zz^{(\mu)} \times \zz^{(\tau)}$ 
is generated by the $\zz$ symmetries of equation~(\ref{Z2}),
\item $H = U(1)_{L_e} \times U(1)_{L_\mu} \times U(1)_{L_\tau}$
is generated by the family lepton-number symmetries of equation~(\ref{U1})
and
\item the permutation group $S_3$ is generated by
the cyclic permutation $C_{e\mu\tau}$ of equation~(\ref{Cemutau})
and the transposition $I_{\mu\tau}$ of equation~(\ref{Imutau}).
\end{description}
The semi-direct product is non-trivial since neither the $\zz^{(\alpha)}$
nor the $U(1)_{L_\alpha}$ commute with $C_{e\mu\tau}$ and $I_{\mu\tau}$.
The elements of $G$ can be written as triples $\left( n, h, s \right)$,
where $n \in N$,
$h \in H$ and $s \in S_3$.
The multiplication law of $G$ is the usual one for semidirect products:
\be
\label{law}
\left( n_1, h_1, s_1 \right) \left( n_2, h_2, s_2 \right)
= \left( n_1 s_1 n_2 s_1^{-1}, h_1 s_1 h_2 s_1^{-1}, s_1 s_2 \right).
\ee
In terms of $3 \times 3$ matrices,
$n$ is represented by a diagonal sign matrix,
$h$ is represented by a diagonal phase matrix
and $s$ is a matrix in the defining triplet representation of $S_3$.
According to section~\ref{fs},
the representations of $G$ that we employ in our model are
\be
\label{list}
\begin{array}{rl}
1 & \mathrm{for} \ \phi_0, \\
n s & \mathrm{for} \ \left( \phi_e, \phi_\mu, \phi_\tau \right), \\
n h s & \mathrm{for} \ \left( e_R, \mu_R, \tau_R \right), \\
h s & \mathrm{for} \ \left( D_{eL}, D_{\mu L}, D_{\tau L} \right)
\ \mbox{and} \ \left( \nu_{e R}, \nu_{\mu R}, \nu_{\tau R} \right), \\
D_2(s) & \mathrm{for} \
\left( \nu_{1R}, \nu_{2R} \right)
\ \mbox{and} \ \left( \chi, \chi^\ast \right),
\end{array}
\ee
where the two-dimensional irreducible represention (irrep) of $S_3$
is denoted $D_2(s)$.
It is easy to convince oneself that 
all the multiplets in the list~(\ref{list}) constitute irreps of $G$.

The group $G$ contains all the family symmetries
of the dimension-four terms of the Lagrangian.
As discussed in detail in section~\ref{model},
there is a sequence of soft-breaking steps
which can be described as
\be
G \stackrel{\dim 3}{\longrightarrow} N \rtimes S_3 
\stackrel{\dim 2}{\longrightarrow} \zz^{(\mu\tau)},
\ee
where $\zz^{(\mu\tau)}$ is the $\zz$ group generated by $I_{\mu\tau}$.

The variations on the symmetries in the following paragraphs
will only concern the normal subgroup $H$ of $G$.

\paragraph{The symmetry group \boldmath$\Delta(6\infty^2)$:}
If we remove 
from the three $U(1)_{L_\alpha}$ the global $U(1)_L$
associated with the total lepton number $L = L_e + L_\mu + L_\tau$,
then the normal subgroup $H$ of $G$ reduces to the set of matrices
\be
\label{Ubg}
U \left( \beta, \gamma \right) = \mbox{diag} \left(
e^{i\beta}, e^{i\gamma}, e^{- i \beta - i \gamma}
\right),
\quad \beta, \gamma \in \left[ 0, 2 \pi \right[.
\ee
In this case,
$H \rtimes S_3$ is the group $\Delta(6\infty^2)$,
or rather a faithful irrep
thereof---see~\cite{fairbairn} for a study of this group.
Therefore,
$G = N \rtimes \Delta(6\infty^2)$.

\paragraph{Switching to \boldmath$\Delta(54)$:}
$\Delta(54)$ is the group
$\Delta(6r^2)$ with $r=3$---for details
see~\cite{fairbairn,luhn,ishimori}.\footnote{The latter paper
uses $\Delta(54)$ for the construction of a lepton flavour model which is,
however,
totally different from ours.}
In this variant of our model we do not use the symmetries $U(1)_{L_\alpha}$.
Instead,
we define the matrix~\cite{trimax}
\be
\label{T}
T \equiv \mbox{diag} \left( 1, \omega, \omega^2 \right),
\ee
and use a symmetry
under which the multiplets transform according to table~\ref{Ttable}.  
\begin{table} \begin{center} \begin{tabular}{c|ccc|c}
& $D_{\alpha L}$ & $\alpha_R$ & $\nu_{\alpha R}$ & $\phi_\alpha$ \\ \hline
$T$ & $T$ & $T^\ast$ & $T$ & $T^2$
\end{tabular} \end{center}
\caption{Transformation of the multiplets under the symmetry $T$.
The multiplets not shown in the table transform trivially.
\label{Ttable}}
\end{table}
The transformation $T$,
together with the $3 \times 3$ permutation matrices,
generates a three-dimensional irrep of $\Delta(54)$.
Notice that this group is \textit{a priori}
smaller---hence less powerful---than $\Delta(6\infty^2)$,
but we enhance its power by allowing it to act non-trivially
on the $\phi_\alpha$. 
It is easy to check that the Yukawa Lagrangian of equation~(\ref{yuwa})
is invariant under $T$,
but we still need the symmetries $\mathbbm{Z}_2^{(\alpha)}$
to remove from $\mathcal{L}_\mathrm{Yukawa}$
possible non-flavour-diagonal terms~\cite{trimax}.
So the symmetry group of our model
is now $G = N \rtimes \Delta(54)$,
which is \emph{finite} and has $8 \times 54 = 432$ elements.
We may still describe $G$ through equation~(\ref{G}),
with $H$ replaced by
\be
H = \left\{ \mbox{diag} \left( \omega^p, \omega^q, \omega^{-p-q} \right)
\ | \ p,q = 0, 1, 2 \right\}.
\ee
Concerning the irreps,
instead of list~(\ref{list}) we now have
\be
\label{list2}
\begin{array}{rl}
1 & \mathrm{for} \ \phi_0, \\
n h^2 s & \mathrm{for} \ \left( \phi_e, \phi_\mu, \phi_\tau \right), \\
n h^\ast s & \mathrm{for} \ \left( e_R, \mu_R, \tau_R \right), \\
h s & \mathrm{for} \ \left( D_{eL}, D_{\mu L}, D_{\tau L} \right)
\ \mbox{and} \ \left( \nu_{e R}, \nu_{\mu R}, \nu_{\tau R} \right), \\
D_2(s) & \mathrm{for} \
\left( \nu_{1R}, \nu_{2R} \right)
\ \mbox{and} \ \left( \chi, \chi^\ast \right).
\end{array}
\ee
The breaking of $T$ is assumed to be soft,
through dimension-three and dimension-two terms.
An important difference relative to section~\ref{model}
is that $T$ removes some of the dimension-four terms
from the scalar potential,
because it acts non-trivially on the $\phi_\alpha$;
one obtains a restricted version of equation~(\ref{V}),
\textit{viz.}
\be
\label{=0}
\lambda_7 = \lambda_8 = 0.
\ee
Notice that,
although we did not use the $U_{L_\alpha}$
in building this variant of the model,
eventually the $U_{L_\alpha}$ turn out to be
(so-called accidental)
symmetries of all the dimension-four terms in the Lagrangian.

\paragraph{Switching to \boldmath$\Delta(6r^2)$
with $r \geq 4$:}
If the $T$ of the previous paragraph is replaced by
\be
\label{Ts}
T = \mbox{diag} \left( 1, \sigma, \sigma^\ast \right),
\quad \sigma \equiv \exp{\left( 2 i \pi / r \right)},
\quad r \geq 4,
\ee
then
\be
H = \left\{ \mbox{diag} \left( \sigma^p, \sigma^q, \sigma^{-p-q} \right)
\ | \ p,q = 0, \dots, r-1 \right\}
\ee
and $H \rtimes S_3$ is isomorphic to $\Delta (6 r^2)$ with $r \geq 4$.
All the previous remarks,
including table~\ref{Ttable},
still hold in this case,
but there is a noteworthy exception:
now we do not need to impose the symmetries $\mathbbm{Z}_2^{(\alpha)}$,
which become just \emph{accidental} symmetries
of all the terms in the Lagrangian with dimension larger than two.
Eventually,
the family symmetry group of the model is again 
of the form of equation~(\ref{G}), 
with $G \cong N \rtimes \Delta (6 r^2)$
having $48 r^2$ elements.

\paragraph{Renormalization-group evolution of \boldmath$\mnu$:}
We proceed to the study of the RGE of $\mnu$
from the seesaw scale down to the electroweak scale.
We first note that the two real degrees of freedom
of the scalar gauge singlet $\chi$ are assumed to be heavy.
Therefore,
the renormalization-group (RG) equations
relevant for the determination of $\mnu$ at the low scale
are simply those of a multi-Higgs-doublet SM.
Those equations were derived in~\cite{renorm}.
It was shown in~\cite{tricp} that
the form of the Yukawa couplings of the charged-lepton fields---see
line~(\ref{alphar})---remains unchanged;
only the value of $y_1$ evolves with the energy scale.
In the same paper~\cite{tricp},
the importance of the quartic scalar couplings
for the RGE of $\mnu$ was investigated;
the following sufficient conditions for RG invariance of $\mnu$ were found:
\begin{enumerate}
\renewcommand{\labelenumi}{\roman{enumi})}
\item The Higgs doublet $\phi_0$,
whose VEV $v_0$ is responsible for generating $\mnu$ at the seesaw scale,
has no Yukawa couplings to the $\alpha_R$.
In our model,
the Yukawa couplings of the charged leptons
are given by line~(\ref{alphar}) at any energy scale.
\item There is a symmetry,
holding at the seesaw scale,
which forbids dimension-five neutrino mass operators
involving two different Higgs doublets.
In our model,
that symmetry is constituted by the three $\mathbbm{Z}_2^{(\alpha)}$.
\item At the seesaw scale there is a symmetry
forbidding quartic couplings of the type
$\left( \phi_k^\dagger \phi_{k^\prime} \right)^2$ $(k \neq k^\prime)$
in the scalar potential.
In our model,
this is satisfied if some symmetry like $T$
leads to the condition~(\ref{=0}).
\end{enumerate}
Thus,
applying the results of our previous paper~\cite{tricp} to the present model,
we find that,
if equation~(\ref{=0}) holds,
then \emph{tri-bimaximal mixing holds at all energy scales
in between the seesaw and electroweak scales}.
According to the preceding discussion,
this is possible by using any of the symmetry groups
$\Delta(6r^2)$ ($r \geq 3$).
On the other hand,
using $\Delta(6\infty^2)$ allows both $\lambda_7$ and $\lambda_8$
to be non-vanishing,
and then corrections to tri-bimaximal mixing from the RGE of $\mnu$
are expected.
Still,
it is well known that such corrections can only be sizable
for a quasi-degenerate neutrino mass spectrum~\cite{balaji}, 
an observation corroborated by explicit studies
of multi-Higgs doublet models~\cite{renorm}
and general considerations~\cite{tanimoto}.

\paragraph{\boldmath$S_3$ versus $S_4$:}
In a series of papers~\cite{lam}
it has been argued that the only finite group
capable of yielding tri-bimaximal mixing is $S_4$,
or else a larger group containing $S_4$.
We want to make some comments on that claim.
Since $S_4 \equiv \Delta(24)$~\cite{luhn},
we can expect that a construction of our model
in analogy to the usage
of $\Delta(6r^2)$ with $r \geq 3$ is possible. 
This is indeed the case.  
We can place the $D_{\alpha L}$,
the $\alpha_R$ and the $\nu_{\alpha R}$ in triplets of $S_4$.
Putting the $\phi_\alpha$
in the \emph{reducible} triplet
representation of the
subgroup $S_3$ and 
adding to this scheme the symmetries $\mathbbm{Z}_2^{(\alpha)}$
in order to avoid non-flavour-diagonal couplings
in $\mathcal{L}_\mathrm{Yukawa}$, 
we can proceed with the construction of the model
just as in section~\ref{model}. 
Actually,
it is easy to see that this way of constructing the model
amounts simply to the replacement of
the $U(1)_{L_\alpha}$ by discrete lepton numbers:
fermions with flavour $\alpha$ are multiplied by $-1$
instead of being multiplied by an arbitrary phase factor.
In the language of equation~(\ref{G}),
in this case the family symmetry group is
$G = (N \times N) \rtimes S_3$---for a complete discussion of its irreps
see~\cite{GL07}.
However,
it appears to us that $S_4$ is not an adequate symmetry group
for our model for two reasons. 
First,
the full symmetry group,
which is only effective in terms of dimension four in the Lagrangian,
is much larger than $S_4$ because its subgroup $S_3$
does not commute with the $\mathbbm{Z}_2^{(\alpha)}$;
therefore,
$S_4$ misses an essential part of the symmetry structure of our model.
Second,
in the terms of dimension three,
\textit{i.e.}~in $\mathcal{L}_\mathrm{Majorana}$,
which are crucial for our model,
the symmetry group is only $S_3$,
something that we had already advocated
in~\cite{prescient}.
In summary,
in our model there is no compelling connection between $S_4$
and tri-bimaximal mixing.

\section{Conclusions}
\label{concl}

In this paper we have proposed a model for tri-bimaximal mixing
based on an extension of the SM with seesaw mechanism and family symmetries.
The scalar sector consists of four Higgs doublets
and one complex gauge singlet,
while the fermion sector has,
besides the SM multiplets,
five right-handed neutrino singlets.
The mixing matrix obtained at the seesaw scale is exactly tri-bimaximal.
The most straightforward version of the model
uses as family symmetries the permutation group $S_3$
together with three $\zz$ symmetries and family lepton numbers;
the latter are softly broken at the seesaw scale.
A slightly more complicated way to obtain the model
makes use of a group $\Delta(6r^2)$ with $r \geq 3$.
The most intricate part of the model
is the stepwise soft symmetry breaking,
which we have tried to explain carefully in section~\ref{model}.
Whether one uses $S_3$ together with family lepton numbers
or a group $\Delta(6r^2)$ does not make any difference,
except for two terms of dimension four in the scalar potential. 
With $\Delta(6r^2)$ those two terms are forbidden and,
as a consequence,
in the one-loop renormalization-group evolution
of the neutrino mass matrix
from the seesaw scale down to the electroweak scale, 
that matrix retains its form
and tri-bimaximal mixing remains exact at the electroweak scale.
With $S_3$ together with family lepton numbers
there are the usual RGE corrections,
which are quite small,
however,
whenever the neutrino mass spectrum is sufficiently non-degenerate.

The main purpose of the model presented here
is to show that in enforcing tri-bimaximal mixing
one does not necessarily require VEV alignment,
supersymmetry,
non-renormalizable terms,
or extra dimensions.
As a further bonus,
one can also obtain RG stability of HPS mixing. 

Finally,
we want to stress that in our model
there is decoupling of the mixing problem from the mass problem;
the latter remains unsolved,
since all lepton masses are completely free.

\paragraph{Acknowledgements:}
We acknowledge support from the European Union
through the network programme MRTN-CT-2006-035505.
The work of L.L.~was supported by the Portuguese
\textit{Funda\c c\~ao para a Ci\^encia e a Tecnologia}
through the project U777--Plurianual.

\newpage

\appendix

\section{Inverting $M_R$}
\setcounter{equation}{0}
\renewcommand{\theequation}{A\arabic{equation}}

The $5 \times 5$ symmetric matrix
\be
M = \left( \begin{array}{ccccc}
M_0 & M_1 & M_1 & M_2 & M_3 \\
M_1 & M_0 & M_1 & \omega^2 M_2 & \omega M_3 \\
M_1 & M_1 & M_0 & \omega M_2 & \omega^2 M_3 \\
M_2 & \omega^2 M_2 & \omega M_2 & M_N & M_4 \\
M_3 & \omega M_3 & \omega^2 M_3 & M_4 & M^\prime_N 
\end{array} \right),
\quad \omega \equiv \exp{\left( 2 i \pi / 3 \right)}
\ee
has non-zero determinant:
\be
\det{M} = \left( M_0 + 2 M_1 \right) \left\{
\left( M_0 - M_1 \right)^2 M_N M^\prime_N
- \left[ \left( M_0 - M_1 \right) M_4 - 3 M_2 M_3 \right]^2 \right\}.
\ee
Let us write
\be
M^{-1} = \left( \begin{array}{cc}
P & R \\ R^T & Q
\end{array} \right),
\ee
where $R$ is a $3 \times 2$ matrix
and $Q$ is a $2 \times 2$ symmetric matrix.
Then,
\be
P = \left( \begin{array}{ccc}
x + y + t &
z + \omega^2 y + \omega t &
z + \omega y + \omega^2 t \\
z + \omega^2 y + \omega t &
x + \omega y + \omega^2 t &
z + y + t \\
z + \omega y + \omega^2 t &
z + y + t &
x + \omega^2 y + \omega t
\end{array} \right),
\ee
with
\ba
x &=& \frac{\left( M_0^2 - M_1^2 \right) \left( M_N M^\prime_N - M_4^2 \right)
+ \left( 4 M_0 + 2 M_1 \right) M_2 M_3 M_4
- 3 M_2^2 M_3^2}{\det{M}},
\\
z &=& \frac{\left( M_1^2 - M_0 M_1 \right)
\left( M_N M^\prime_N - M_4^2 \right)
+ \left( M_0 - 4 M_1 \right) M_2 M_3 M_4
- 3 M_2^2 M_3^2}{\det{M}},
\\
y &=& \frac{\left( M_0 + 2 M_1 \right) M_2^2 M^\prime_N}{\det{M}},
\label{y} \\
t &=& \frac{\left( M_0 + 2 M_1 \right) M_3^2 M_N}{\det{M}}.
\label{t} 
\ea

\end{document}